\documentclass[pdflatex,sn-mathphys-num]{sn-jnl}

\usepackage{graphicx}%
\usepackage{multirow}%
\usepackage{amsmath,amssymb,amsfonts}%
\usepackage{amsthm}%
\usepackage{mathrsfs}%
\usepackage[title]{appendix}%
\usepackage{xcolor}%
\usepackage{textcomp}%
\usepackage{manyfoot}%
\usepackage{booktabs}%
\usepackage{algorithm}%
\usepackage{algorithmicx}%
\usepackage{algpseudocode}%
\usepackage{listings}%
\usepackage{tikz}

\theoremstyle{thmstyleone}%
\theoremstyle{thmstyletwo}%

\theoremstyle{thmstylethree}%

\begin{document}

\title[Persistent Homology as a Morphological Signature of Fibrin Networks]{Persistent Homology as a Morphological Signature of Fibrin Networks}


\author[1]{\fnm{Thomas} \sur{Burnett}}\email{thomas.burnett@anu.edu.au}
\equalcont{These authors contributed equally to this work.}

\author[2]{\fnm{Theresa} \sur{Reinhold}}\email{trkn20@student.aau.dk}
\equalcont{These authors contributed equally to this work.}

\author[3]{\fnm{Bea} \sur{Bleile}}\email{bbleile@une.edu.au}

\author[4]{\fnm{Sophie} \sur{Raynor}}\email{sophie.raynor@jcu.edu.au}

\author[5]{\fnm{Freya} \sur{Jensen}}\email{freya.jensen@iwr.uni-heidelberg.de}


\author[6]{\fnm{Martin} \sur{Hermann}}\email{martin.hermann@tirol-kliniken.at}

\author[7,8]{\fnm{Tua} \sur{Gyldenholm}}\email{tuagyl@clin.au.dk}

\author*[9,10]{\fnm{Yossi} \sur{Bokor Bleile}}\email{yossi.bokorbleile@ist.ac.at}

\affil[1]{\orgdiv{Mathematical Sciences Institute}, \orgname{Australian National University}, \city{Canberra}, \country{Australia}}

\affil[2]{\orgdiv{Department of Mathematical Sciences}, \orgname{Aalborg University}, \city{Aalborg}, \country{Denmark}}

\affil[3]{\orgdiv{School of Science and Technology}, \orgname{University of New England}, \city{Armidale}, \country{Australia}}

\affil[4]{\orgdiv{College of Science and Engineering}, \orgname{James Cook University}, \city{Townsville}, \country{Australia}}

\affil[5]{\orgdiv{Interdisciplinary Center for Scientific Computing}, \orgname{Heidelberg University}, \city{Heidelberg}, \country{Germany}}

\affil[6]{\orgdiv{Department of General and Surgical Intensive Care Medicine}, \orgname{Medical University of Innsbruck}, \city{Innsbruck}, \country{Austria}}

\affil[7]{\orgdiv{Department of Clinical Medicine}, \orgname{Aarhus University}, \city{Aarhus}, \country{Denmark}}

\affil[8]{\orgdiv{Department of Clinical Biochemistry}, \orgname{Aarhus University Hospital}, \city{Aarhus}, \country{Denmark}}

\affil*[9]{\orgname{Institute of Science and Technology Austria}, \city{Klosterneuburg}, \country{Austria}}

\affil[10]{\orgdiv{School of Mathematics and Statistics}, \orgname{University of Sydney}, \city{Sydney}, \country{Australia}}

\abstract{
    We present an investigation of the applicability of Topological Data Analysis (TDA) to the study of high-resolution confocal microscopy images of fibrin network structures from patients with \textcolor{black}{esophageal} cancer undergoing intended curative surgery. Investigation of clot structure \textcolor{black}{at various points in time} brings new knowledge about blood coagulation, risk of bleeding, and thrombosis in this group of patients. Images of fibrin network formation in the collected blood samples were captured by confocal microscopy and three-dimensional z-stacks were \textcolor{black}{analyzed}. Each z-stack was cropped to a \textcolor{black}{center} region for analysis, the validity of which is assessed in detail. \textcolor{black}{Overall, we found no consistent differences in fibrin network structures between standard and intervention groups, as well as no significant differences in network topology across the perioperative period.}
}



\maketitle

\section{Introduction}
    Fibrin protein networks play a crucial role in hemostasis and thrombosis by providing the structural framework for blood coagulation. Fibrin strands are formed when fibrinogen is cleaved by thrombin. The resulting fibrin monomers \textcolor{black}{polymerize}, adhering to each other, and form a three-dimensional network that \textcolor{black}{stabilizes} the platelets that make up the otherwise fragile, primary clot. The fibrin network is eventually cleaved by plasmin and dissolved; a process known as fibrinolysis \cite{Weisel2005}.

    The \textcolor{black}{organization and} architecture of fibrin networks are believed to influence clot stability. The tightness and density of the networks might influence mechanical properties of the clot, as well as the efficacy and speed of fibrinolysis and therefore clot longevity \cite{Zabczyk2024}. Several methods have been developed for use in research to investigate fibrin networks, but quantifying the complexity of fibrin networks remains challenging due to their heterogeneous, three-dimensional nature \cite{Undas2024}. Confocal microscopy has recently provided valuable insights into fibrin structure \cite{Fenger-Eriksen2020}.

    Analysis of coagulation activity forms an important part in assessing and treating both bleeding and thrombosis in at-risk patient groups. It is well established that cancer impacts the normal physiological balance of clot formation and dissolution, increasing both the risk of bleeding and thrombosis \cite{Papa2007}. Much research investigating initiation of coagulation and clot formation has been performed, and several well established analyses applicable in \textcolor{black}{research} and clinical use exist. However, there are few analyses that investigate fibrinolysis \cite{Undas2024}. The geometry of fibrin networks may play an important role in the stability and durability of the formed blood clot, but few studies have attempted to investigate this \cite{Weisel2005}. Furthermore, knowledge of fibrin network structure in different clinical presentations, including patients treated with anticoagulation, remains poorly \textcolor{black}{characterized}.

    Topological Data Analysis (TDA) offers a novel approach for \textcolor{black}{characterizing} fibrin architecture by capturing structural features that persist across multiple spatial scales. \textcolor{black}{Topological features of patient-level imaging data have been shown to predict clinical outcomes in glioblastoma \cite{Crawford2020}.} Unlike conventional morphological metrics - such as fiber length, branching density, and porosity - TDA provides a robust, quantitative framework for assessing global and local network properties through tools like persistent homology. These methods enable the measurement of topological features that have been shown to be feasible for detecting dilution coagulopathy in a porcine model \cite{Berger2017}\textcolor{black}{, and differences between human and monkey fibrin networks in scanning electron microscopy images~\cite{Berry2020}}, which suggest their potential applicability to the study of clot stability, disease states, or therapeutic interventions.

    In \cite{Berger2017, Berger2024}, the authors used persistent homology on 3D \textcolor{black}{grayscale} images of fibrin networks\textcolor{black}{, averaged over the z-coordinate to two-dimensional images,} in porcine blood samples to understand the effects of dilution on the persistent homology of the networks. This method was then \textcolor{black}{successfully} used to investigate clot characteristics in four patients as part of a study investigating clot properties and coagulation in children undergoing major craniofacial surgery. In \textcolor{black}{their} study, the two patients who received tranexamic acid, a drug which inhibits the \textcolor{black}{degradation} of fibrin, had denser networks than the two patients who received placebo \cite{Fenger-Eriksen2020}. While the results of this study naturally should be interpreted with caution due to the small cohort, it adds merit to the idea that \textcolor{black}{characterization} of fibrin networks might be a useful addition to established coagulation analyses. The present study aims to further investigate the applicability of TDA to the analysis of fibrin networks.

    In this study, we use \textcolor{black}{zero-, one- and two-dimensional} persistent homology to \textcolor{black}{analyze} images of fibrin networks obtained from human blood samples. Furthermore, we compare the possible effects of two different lengths of post-surgical anticoagulant regimes on fibrin network formation, \textcolor{black}{as well as study the topology of the networks at four different sample times}. By \textcolor{black}{analyzing} the topological characteristics of fibrin structures in patients receiving different durations of anticoagulant prophylaxis, we aim to identify differences in \textcolor{black}{network} architecture that might reflect dysregulation of hemostasis. This approach provides a quantitative framework for investigating fibrin networks in human blood samples.

    The \textcolor{black}{main} contributions of this study are: \textcolor{black}{1) a modified methodology which computes persistent homology of the full three-dimensional z-stacks, rather than of a two-dimensional image obtained by averaging over the z-coordinate as in previous work~\cite{Berger2017,Berger2024}, and so allows for the 3-dimensional structure of the fibrin network, including voids enclosed by the network, to be quantified, 2) the application of persistent homology to a longitudinal study to compare the effects of two different post-surgical anticoagulant regimes on fibrin network formation, }3) an investigation of the homogeneity of the fibrin network within individual \textcolor{black}{z-stacks, to assess how representative a single cropped region is of the whole field of view}, and 4) the introduction of additional existing methods within TDA to the study of fibrin\textcolor{black}{: PERMANOVA based on the $2$-Wasserstein distances, and clustering of four persistence signatures (Betti curves, persistence landscapes, persistence images and rank functions). Despite the pipeline not detecting significant morphological differences between the regimes, it is a controlled negative result that establishes the method’s baseline behavior on clinical samples. Thus it is a first, necessary step before the clinically motivated comparisons, healthy versus pathological, that the framework makes possible but that remain to be performed.}

\section{Materials and methods}\label{sec:materials-methods}
    The \textcolor{black}{blood} used for the present study were collected as part of the Thromboprophylaxis in Oesophageal Cancer Patients Randomised Controlled Trial (TOP-RCT) study \cite{Gyldenholm2024}. Adult patients with \textcolor{black}{esophageal} cancer referred for intended curative surgery at Aarhus University Hospital, Aarhus, Denmark, were included between September 2021 and April 2024. Patients were randomly assigned to receive either 10 or 30 days of prophylactic anticoagulant treatment with the low molecular weight heparin Fragmin (Pfizer, USA) aimed at managing coagulation and thrombosis risk. All patients received 5000 International Units (IU) daily in the form of an injection. The standard group received prophylaxis until discharge from the hospital, and the intervention group received prophylaxis for one month after surgery. Blood \textcolor{black}{was} drawn before, immediately after surgery, one day post-surgery, and thirty days post-surgery. \textcolor{black}{Ten patients}, five in each treatment group, \textcolor{black}{were selected to be part of our study}. \textcolor{black}{All blood draws were centrifuged for 25 minutes at 3000 g and stored at -80 degrees Celsius until analysis.}

\subsection{\textcolor{black}{Data curation methodology}}

    For imaging, \textcolor{black}{drawn blood was} thawed and incubated for 30 minutes at room temperature with addition of the commercially available reagents StarTem and ExTem (Vingmed, Denmark) to induce coagulation and fibrin network formation. Furthermore, a solution containing a fluorescein isothiocyanate-marked fibrin-binding peptide developed and described in \cite{Weiss2017} was added to \textcolor{black}{visualize} the fibrin formed in the \textcolor{black}{blood}.
    
    Confocal microscopy was performed with a spinning disc confocal system (UltraVIEW VoX, Perkin Elmer, USA) with a Zeiss AxioObserver Z1 microscope (Zeiss, Germany). Images were obtained using Volocity software (Quorum Technologies Inc., Canada) with a 40x water immersive objective. This resulted in image stacks with $1000 \times 1000$ pixels, $1$ $\mu$m pixel size, and $0.5$ $\mu$m Z-steps.
    
\subsection{Image pre-processing}
\textcolor{black}{Illumination is strongest in the center of the field of view, so we restrict the analysis to a cropped central window.} For our 3D pipeline, we compute the center of the data by finding the median of the cumulative intensity distribution along each \textcolor{black}{lateral} axis, and then crop the $x$ and $y$ axes to obtain a $400 \times 400$ \textcolor{black}{pixel} window \textcolor{black}{(see Section~\ref{sec:robustness-cropping})}. The z-stack is then smoothed and downsampled along the axial direction by averaging neighboring slices with a 3-slice box filter and subsampling with stride two\textcolor{black}{, giving an image of $11\times 400 \times 400$ voxels}. A mild Gaussian filter ($\sigma = 1$ \textcolor{black}{pixel}) is applied in the $x$ and $y$ dimensions only. Intensity values are \textcolor{black}{normalized} to the $[0,1]$ range by clipping to the 1st and 99th percentiles. Finally, the image is inverted so that fibrin \textcolor{black}{fibers} appear dark against a bright background \textcolor{black}{so that they enter the sublevel set filtration described below first}.

\subsection{Topological data analysis}

Persistent homology is a tool of Topological Data Analysis (TDA) which captures robust and explainable quantitative features of data\textcolor{black}{~\cite{Edelsbrunner2010}}. The central idea is that when datasets can be turned into filtrations $\{X_t: t\in\mathbb{R}\}$ (due to our \textcolor{black}{normalization} we can restrict to $t \in [0,1]$), then meaningful information about local and global structure of the data may be captured by the manner in which various topological features (e.g. connected components, cycles, and voids) are formed (born), destroyed (die), or continue to exist (persist) as the filtration parameter is varied.  

When working with images, we often use \emph{cubical complexes} \cite{Kaczynski2004} to represent our data. Just as a simplicial complex is built from vertices, edges, triangles and tetrahedra, a cubical complex is built from vertices, edges, squares and cubes, such that the resulting collection is closed under taking faces. 

In particular, we can think of each \textcolor{black}{voxel as a cube}, and obtain a \textcolor{black}{three-}dimensional cubical complex $X$ by adding in all (possibly shared) \textcolor{black}{squares,} edges and vertices. A \textcolor{black}{grayscale} image, defined by an intensity function on the \textcolor{black}{voxels}, describes a \textcolor{black}{filtration of $X$, known as the \emph{T-construction}~\cite{Bleile2022},} whereby, for each value $t$ of the intensity parameter, $X_t$ is the smallest \textcolor{black}{three-dimensional} subcomplex of $X$ that includes all \textcolor{black}{voxels (cubes)} with \textcolor{black}{grayscale} intensity less than or equal to $t$.

\begin{figure}[htbp]
      \centering
      \includegraphics[width=\linewidth]{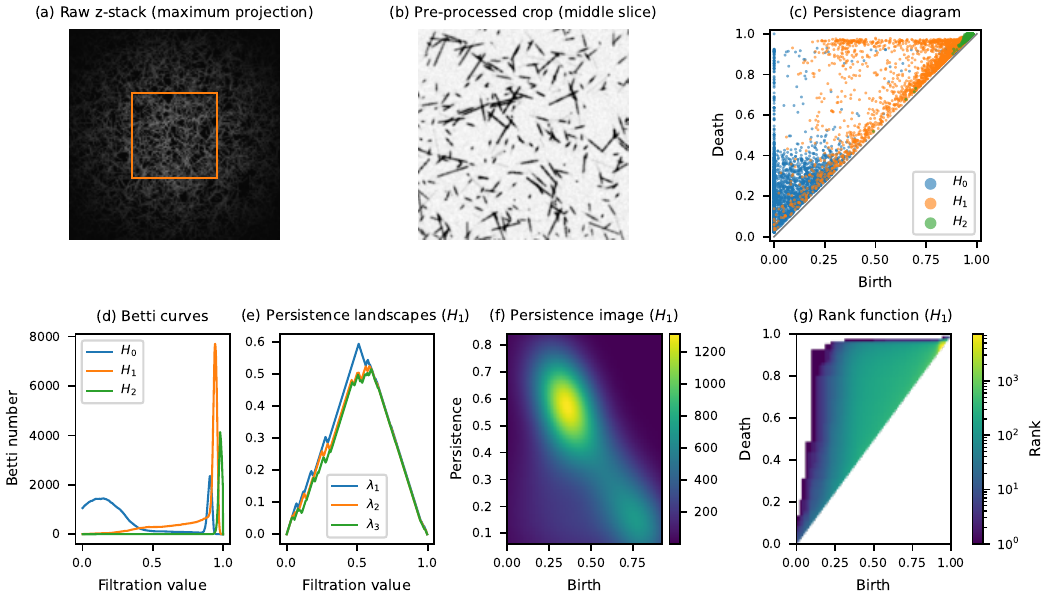}
      \caption{\textcolor{black}{The analysis pipeline illustrated on a synthetic z-stack, generated similarly to the Pickup Sticks Simulator of~\cite{Berry2020}, since the patient images cannot be shown. (a) Maximum projection of the raw z-stack, with $400 \times 400$ pixel center crop. (b) Middle slice of the pre-processed crop. (c) Persistence diagram. (d)--(g) The four persistence signatures used for clustering.}}
      \label{fig:pipeline}
\end{figure}

Thus we have a filtration of the cubical complex corresponding to the \textcolor{black}{$11 \times 400 \times 400$ voxel} image. From this filtered complex, we obtain \textcolor{black}{persistence diagrams in homological degrees $0$, $1$ and $2$, computed with coefficients in $\mathbb{Z}/2\mathbb{Z}$}, which we use as a representation of the structure of the corresponding fibrin network. The GUDHI library \cite{gudhi} was used to create the cubical complexes from the images and to compute the persistence diagrams from these. \textcolor{black}{Fig~\ref{fig:pipeline} illustrates the pipeline on a synthetic z-stack.}

Each point \textcolor{black}{$(b,d)$} in the persistence diagram represents a topological feature, i.e. a connected component, a cycle or a void. The birth value \textcolor{black}{$b$} is the filtration parameter value such that the corresponding feature is formed in the complex \textcolor{black}{$X_b$}. The death value \textcolor{black}{$d$} is the parameter value at which the corresponding feature is destroyed. That is, the feature is present for all values \textcolor{black}{$b\leq t<d$}, but not for \textcolor{black}{$t<b$ or $t\geq d$}. If a feature persists, we set \textcolor{black}{$d = 1$}. Points close to the diagonal in the diagram correspond to features that only \textcolor{black}{exist} for a short period. The further away from the diagonal, the more persistent the corresponding feature. \textcolor{black}{We use the \emph{2-Wasserstein distance} \cite{Skraba2025,PDaMSoC} on persistence diagrams as a dissimilarity score between fibrin networks.}

\textcolor{black}{For two persistence diagrams $D$ and $D'$ of the same homological degree, a \emph{matching} $M$ between $D$ and $D'$ is a subset of $(D \cup \{\Delta\}) \times (D' \cup \{\Delta\})$ in which each point of $D$ and $D'$ appears as many times as its multiplicity, and $\Delta$ is an abstract element representing the diagonal. Points are compared using the Euclidean distance, with the distance from a point $(b,d)$ to the diagonal being the perpendicular distance $\|(b,d) - \Delta\|_2 = |d - b|/\sqrt{2}$, and $\|\Delta - \Delta\|_2 = 0$. The $2$-Wasserstein distance is
\begin{align*}
  W_2(D,D') = \inf_{M} \left( \sum_{(x,y) \in M} \|x - y\|_2^2 \right)^{1/2},
\end{align*}
where the infimum is over all matchings between $D$ and $D'$. For two fibrin networks with diagrams $D_k$ and $D'_k$ in degree $k=0,1,2$, we use the total $2$-Wasserstein distance
\begin{align*}
  W_2^{\mathrm{tot}} = \left( \sum_{k=0}^{2} W_2(D_k, D'_k)^2 \right)^{1/2}.
\end{align*}
Since essential classes are assigned death values of $1$, all diagrams are finite. Before computing distances, we remove all points with persistence $d-b < 0.02$ for computational tractability. Distances are computed with the Hera library~\cite{Kerber2017} in GUDHI, which approximates $W_2$ to within a relative error of $1\%$.
}    
    
There exist several \textcolor{black}{vectorizations} of persistence diagrams that allow for topological information to be \textcolor{black}{analyzed} with \textcolor{black}{statistics or} machine learning.  We convert the persistence diagrams to four different \textcolor{black}{vectorization} representations of the topology: Betti \textcolor{black}{curves, persistence landscapes, persistence images and rank functions} \cite{Ali2023,Robins2016}. \textcolor{black}{Each feature is standardized to zero mean and unit variance across the z-stacks, and k}-means clustering is then used to partition the \textcolor{black}{z-stacks} into $k$ distinct groups, where $k$ is determined with elbow plots. \textcolor{black}{When converting to Betti curves we set the axis resolution to $1000$. The first three landscape functions were calculated with a resolution of $200$ and a sampling range defined by the minimum birth and maximum death. For persistence images we set the spread to $0.1$, the resolution to $50$ and sampled the $1000$ points with the highest persistence\textcolor{black}{, weighted linearly by their persistence}. Rank functions were computed \textcolor{black}{in homological degrees $0$ and $1$} with a sampling resolution of $80$.}

\subsection{Statistical \textcolor{black}{a}nalysis}
In comparing different collections of persistence diagrams, we used permutational multivariate analysis of variance (PERMANOVA) \cite{Anderson2001} based on \textcolor{black}{total} $2$-Wasserstein distances. \textcolor{black}{The space of persistence diagrams with $2$-Wasserstein distance is a complete, separable metric space, so probability distributions on it, and their Fr\'echet means, are well defined~\cite{Mileyko2011}; however, it is not a vector space and these Fr\'echet means need not be unique~\cite{Turner2014}, so classical ANOVA methods which compare group means are not directly applicable.} Similar to ANOVA methods, PERMANOVA tests a null hypothesis that the centroids of groups are equivalent through comparing between-group and within-group variation, but \textcolor{black}{it assesses significance} through comparing the \textcolor{black}{pseudo-F statistic to its values under random permutations of the group labels, which are exchangeable under the null hypothesis. This removes} assumptions of normality and \textcolor{black}{the} need for computing averages.

The pseudo F-statistic is computed with
\begin{align*}
    F &= \frac{SS_T-SS_W}{SS_W}\frac{N-p}{p-1}\\
    SS_T &= \frac{1}{N}\sum_{i=1}^{N-1}\sum_{j=\textcolor{black}{i}+1}^{N}d_{ij}^2\\
    SS_W &= \frac{1}{\textcolor{black}{n}}\sum_{i=1}^{N-1}\sum_{j=\textcolor{black}{i}+1}^{N}d_{ij}^2\delta_{i,j}
\end{align*}

where $p$ is the number \textcolor{black}{of} groups, $N=pn$ is the total number of diagrams, $d_{ij}^2$ is the squared $2$-Wasserstein distance between diagrams $i$ and \textcolor{black}{$j$}, and $\delta_{ij}$ is $1$ if diagrams $i$ and $j$ belong to the same group, otherwise $0$. \textcolor{black}{For groups of unequal size, we use the general form of $SS_W$ given in~\cite{Anderson2001}, in which the within-group sum for each group is divided by that group's size.}

The $p$-value is then given by the ratio of the number of permutations where this computed pseudo F statistic is greater than or equal to the non-permutated case, and the total number of permutations. \textcolor{black}{For the exact test, the observed assignment of group labels is one of the enumerated permutations. For a Monte Carlo test, we count the observed assignment as one additional permutation.}
  
\section{Results}

\subsection{\textcolor{black}{Treatment group comparison}}\label{sec:treatment}
\begin{figure}[htbp]
    \centering
    \includegraphics[width=\linewidth]{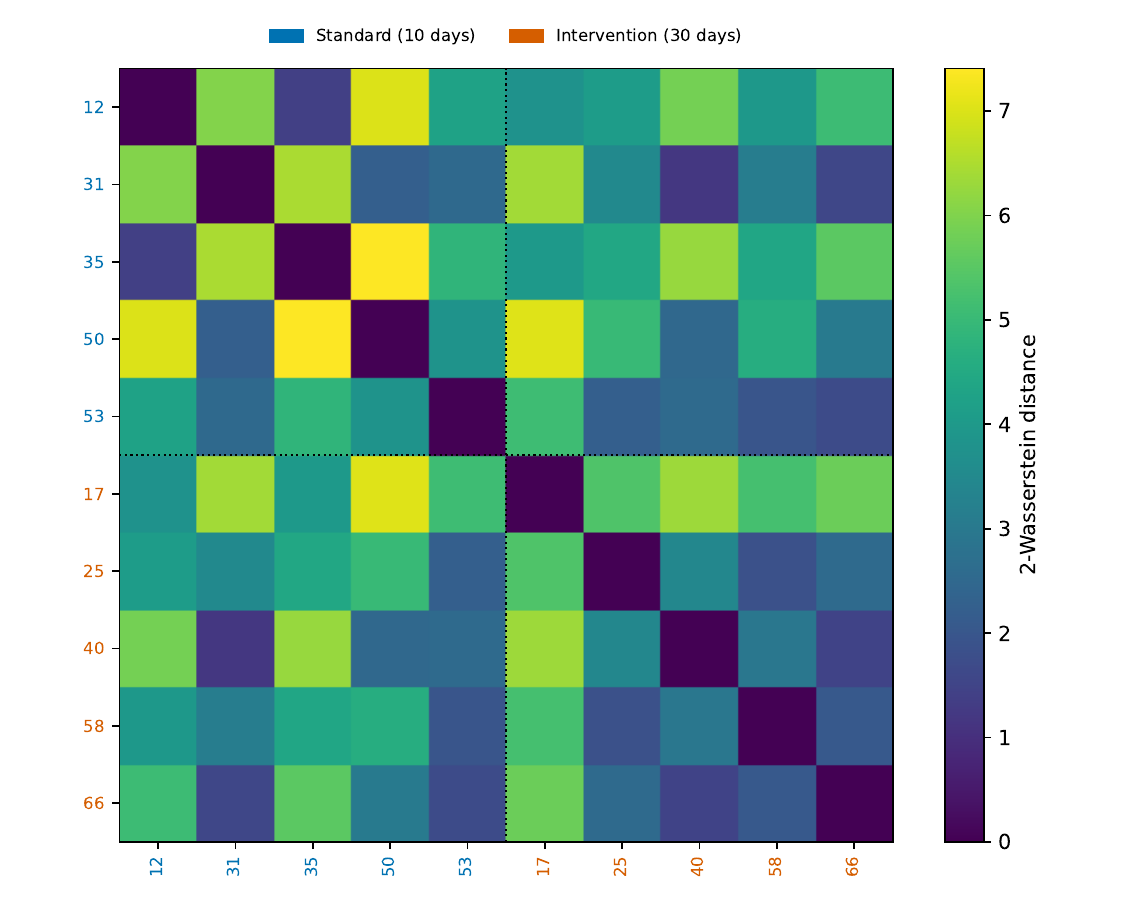}
    \caption{Heat matrix of $2$-Wasserstein distances at 30 days post-surgery.}
    \label{fig:heatmap-treatmentgroups}
\end{figure}

To assess \textcolor{black}{whether the duration of prophylaxis affects the topology of the fibrin network, we compared} the two treatment groups at 30 days post-surgery \textcolor{black}{($p=2$ groups of $n=5$ patients, $N=10$ z-stacks)}. \textcolor{black}{W}e computed pairwise total $2$-Wasserstein distance between persistence diagrams, after filtering \textcolor{black}{the points in the persistence diagrams to points} with minimum persistence $0.02$. The resulting dissimilarity matrix is \textcolor{black}{visualized} in Fig~\ref{fig:heatmap-treatmentgroups}.

\begin{figure}[htbp]
    \centering
    \includegraphics[width=\linewidth]{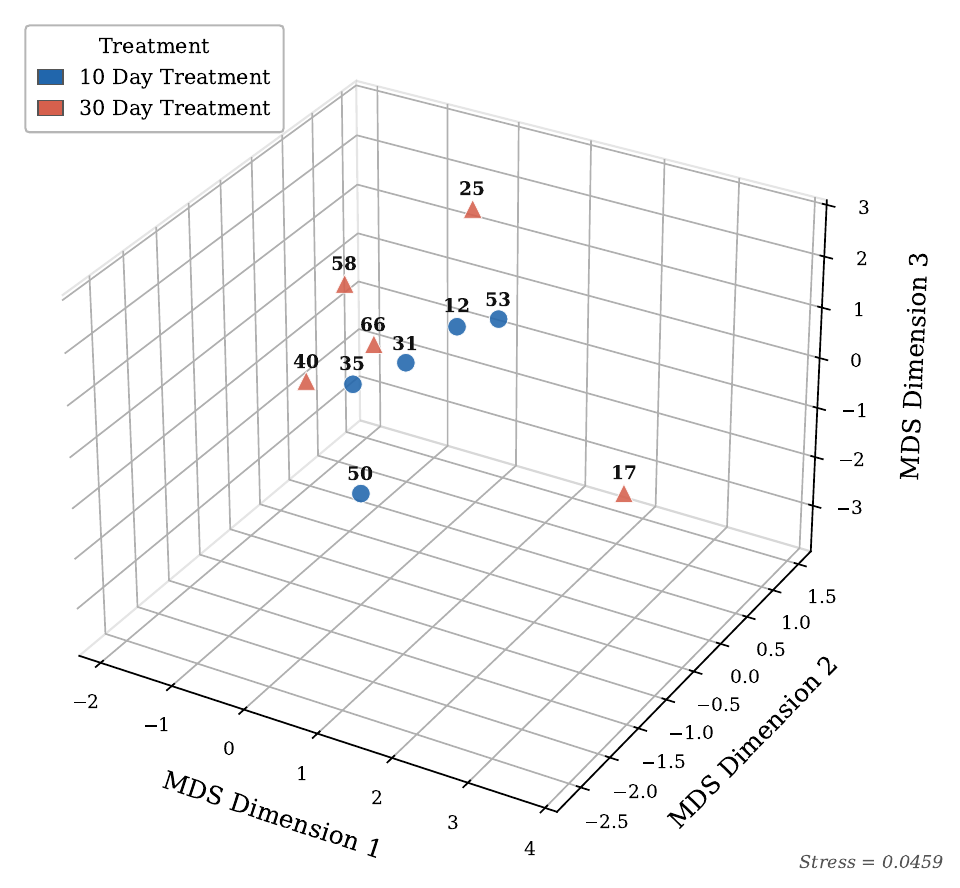}
    \caption{3D MDS of $2$-Wasserstein distances at 30 days post-surgery.}
    \label{fig:mds-treatmentgroups}
\end{figure}

For the PERMANOVA test, all 252 possible \textcolor{black}{assignments of the ten patients to the two groups of five} were enumerated for an exact $p$-value. The observed pseudo-F statistic was \textcolor{black}{$0.3576$}, indicating that between-group variation was less than within-group variation. The exact $p$-value was \textcolor{black}{$0.8571$}, providing no evidence of difference in \textcolor{black}{network topology} between the two treatment groups. However, \textcolor{black}{the} associated multi-dimensional scaling (MDS) plot Fig~\ref{fig:mds-treatmentgroups} (see \cite{PDaMSoC} for a discussion on MDS) \textcolor{black}{shows patient 17 as an outlier. To assess its influence, we repeated the test without patient 17 (pseudo-F $= 0.7916$, exact $p= 0.4444$), and without both patients 17 and 58 (pseudo-F $= 0.6929$, exact $p= 0.4643$). Neither result is significant, and note that with $126$ and $56$ possible assignments respectively, these exact tests are coarse.}

The clustering of the four different \textcolor{black}{vectorizations} of persistence diagrams \textcolor{black}{provides} no evidence for an observable difference between treatment groups. Furthermore, we observe that the \textcolor{black}{patients} cluster more accurately into three \textcolor{black}{or four} clusters. Thus, the two patient groups show no observable topological differences. Similar to the conclusions of the PERMANOVA test, when \textcolor{black}{a patient does} cluster individually\textcolor{black}{, it is most often} patient 17.

\subsection{\textcolor{black}{Pre-operative and post-operative comparison}}
\begin{figure}[htbp]
    \centering
    \includegraphics[width=\linewidth]{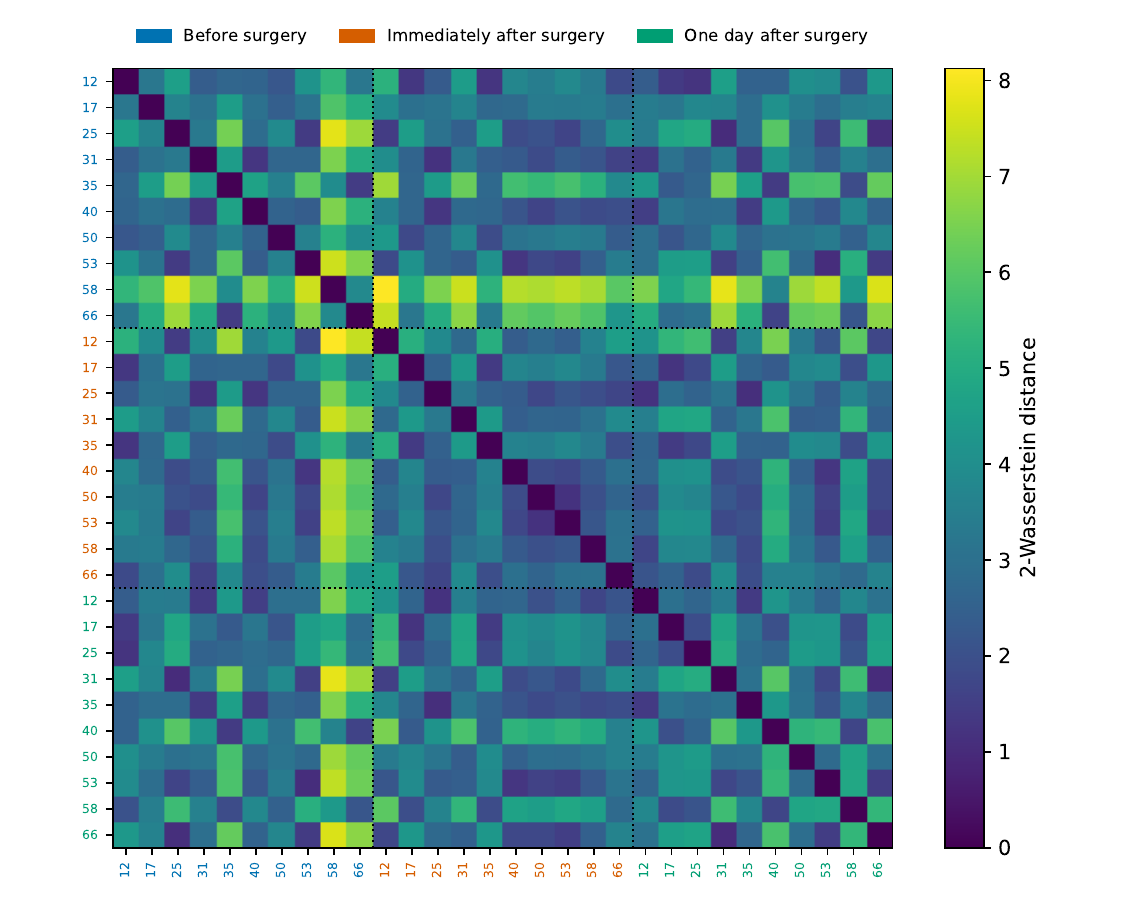}
    \caption{\textcolor{black}{Heat matrix of $2$-Wasserstein distances across three blood draw times.}}
    \label{fig:heatmap-times}
\end{figure}

\begin{figure}[htbp]
    \centering
    \includegraphics[width=\linewidth]{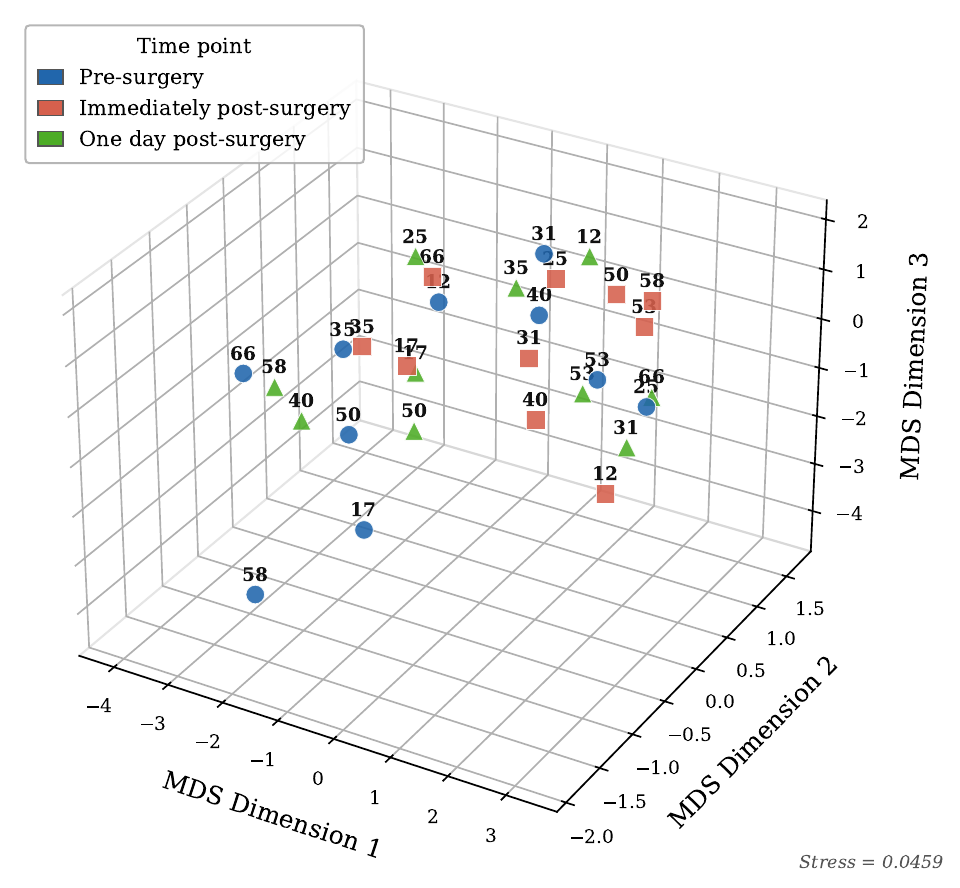}
    \caption{3D MDS of $2$-Wasserstein distances across three \textcolor{black}{blood draw times}.}
    \label{fig:mds-times}
\end{figure}

To assess \textcolor{black}{whether the topology of the fibrin network changes over the perioperative period, we compared} the fibrin networks before surgery, immediately after surgery, and one day post-surgery \textcolor{black}{(three blood draw times of ten patients each, with $30$ z-stacks in total). The 30-day blood draw is not included, since by then the two treatment groups have received different durations of prophylaxis. This is instead analyzed in Section~\ref{sec:treatment}. W}e pairwise computed total $2$-Wasserstein distance between all persistence diagrams, after filtering \textcolor{black}{the points in the persistence diagrams to those with} minimum persistence $0.02$. The resulting dissimilarity matrix is \textcolor{black}{visualized} in \textcolor{black}{Fig~\ref{fig:heatmap-times}, and its MDS embedding in} Fig~\ref{fig:mds-times}.

A one-way PERMANOVA was conducted to assess whether \textcolor{black}{network topology} differed across the three \textcolor{black}{blood draw times}, using 9,999 Monte Carlo permutations\textcolor{black}{, since the number of possible assignments of the 30 z-stacks to three groups of ten (approximately $5.6 \times 10^{12}$) is too large to enumerate}. The observed pseudo-F statistic was \textcolor{black}{$1.2523$ ($p = 0.2644$)}, providing no evidence of an overall difference in \textcolor{black}{network topology} across the \textcolor{black}{blood draw times}. \textcolor{black}{Since each patient contributes one z-stack per blood draw time, we also conducted the test with permutations restricted to within patients, giving pseudo-F $= 1.2523$ ($p = 0.2563$)}. Pairwise \textcolor{black}{PERMANOVA} tests were also conducted for each pair of \textcolor{black}{blood draw times}, with Bonferroni correction applied to control for the multiple \textcolor{black}{comparisons}. The greatest separation was observed between pre-surgery and immediate post-surgery, with pseudo-F statistic \textcolor{black}{$2.3057$ (corrected $p=0.2901$)}, although this did not reach significance. The other two comparisons had corrected \textcolor{black}{$p=1$}. Taken together, these results provide no statistical evidence of a change in composition across the \textcolor{black}{blood draw times}, \textcolor{black}{though} the small \textcolor{black}{cohort} size limits the power of these tests.

The clustering of the four different \textcolor{black}{vectorizations} of persistence diagrams also \textcolor{black}{provides} no evidence for an observable difference between \textcolor{black}{blood draw} times. No single \textcolor{black}{blood draw time} is topologically significant compared to other \textcolor{black}{blood draw times} as no clear pattern of behavior is observable for any single \textcolor{black}{blood draw time}. However, a few \textcolor{black}{blood draws} consistently cluster individually. Upon inspection of these blood draws, it was observed that the corresponding images contained debris not related to the fibrin network. \textcolor{black}{Thus, we found no} topological differences between patient groups; however, it may be capable of detecting outliers with debris in the sample.

\subsection{Robustness of cropping}\label{sec:robustness-cropping}
We conducted three experiments to assess the validity of the center-crop procedure. The first established that topology changes gradually with spatial displacement from the center, the second tested whether the center crop is representative of the image as a whole, and the third assessed whether crop-level variation is \textcolor{black}{negligible} relative to the \textcolor{black}{between-patient} signal of interest.

\begin{figure}[htbp]
    \centering
    \includegraphics[width=\linewidth]{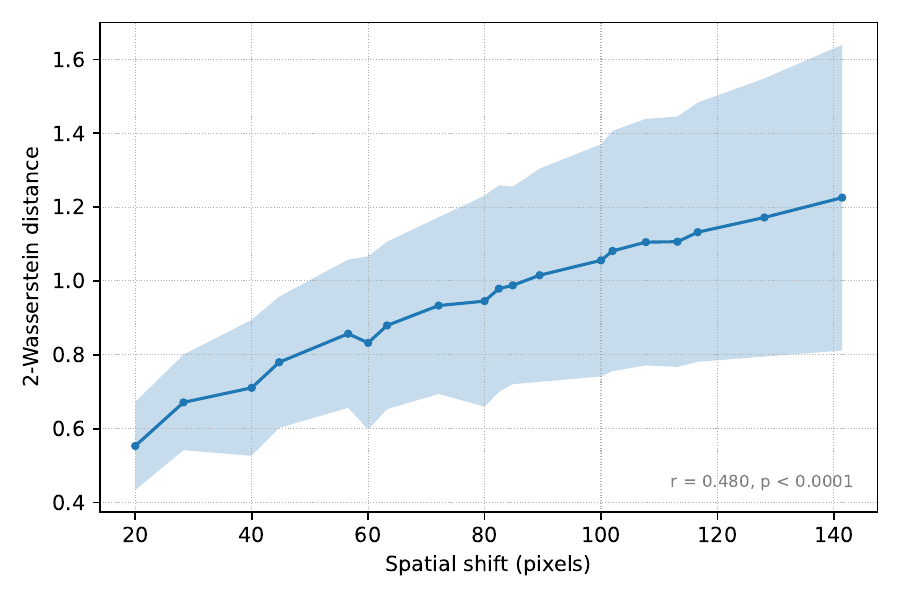}
    \caption{Mean $2$-Wasserstein distance between the center crop and spatially
    shifted crops, as a function of shift magnitude (pixels). Points show the
    mean across all \textcolor{black}{$40$ z-stacks}; shaded region shows $\pm 1$ standard deviation.
    Pearson correlation \textcolor{black}{$r = 0.480$} ($p < 0.0001$).}
    \label{fig:cropping-shift}
\end{figure}

In the first experiment, we computed persistence diagrams for crops centered at positions shifted in increments of $20$ pixels in both $x$ and $y$ directions, ranging from $-100$ to $100$ pixels. \textcolor{black}{For each z-stack, we computed the total} $2$-Wasserstein distance \textcolor{black}{between the diagram of the center crop and the diagram of each of the $120$ shifted crops of the same z-stack. This distance} displayed a \textcolor{black}{tendency} to increase with spatial displacement, from a mean of \textcolor{black}{$0.5532$ (s.d. $0.1192$)} at the smallest shift of $20$ pixels to \textcolor{black}{$1.2256$ (s.d. $0.4139$)} at the largest shift of $141$ pixels, with Pearson correlation \textcolor{black}{between shift magnitude and distance, over all $40 \times 120$ pairs,} of \textcolor{black}{$r=0.480$} ($p<0.0001$); see Fig~\ref{fig:cropping-shift}. The relationship was gradual and consistent (averaged across the $40$ \textcolor{black}{z-stacks}), and indicates that the \textcolor{black}{center} crop is a locally stable summary of the fibrin network topology. This complements the pessimistic theoretical stability guarantees of \cite{Cohen2007} and \cite{Skraba2025} with context specific to our imaging setting.

In the second experiment, for each of the z-stacks, we generated ten random $400\times400$ pixel crops\textcolor{black}{, with centers drawn uniformly from all positions at which the crop lies within the field of view, and then pre-processed them identically to the center crop}. The mean total $2$-Wasserstein distance between the center crop and random crops was \textcolor{black}{$1.6189$ (s.d. $0.5466$)}, compared to a mean between-random crop distance of \textcolor{black}{$1.5215$ (s.d. $0.4245$)}, giving a mean ratio of \textcolor{black}{$1.0595$ (s.d. $0.1875$)}. A one-sample $t$-test against a null ratio of $1$ was \textcolor{black}{not significant ($t=2.008$, $p=0.0517$), so the center crop is at most slightly} more topologically distinct from random crops than random crops are from each other.

\textcolor{black}{The mean ratio slightly above $1$ is consistent with the illumination profile of the microscope, as the random crops extend into the less strongly illuminated periphery, while the center crop is the most evenly illuminated. In any case, comparisons across individuals are unaffected}, since the same region is used consistently for all \textcolor{black}{z-stacks}.

\begin{figure}[htbp]
    \centering
    \includegraphics[width=\linewidth]{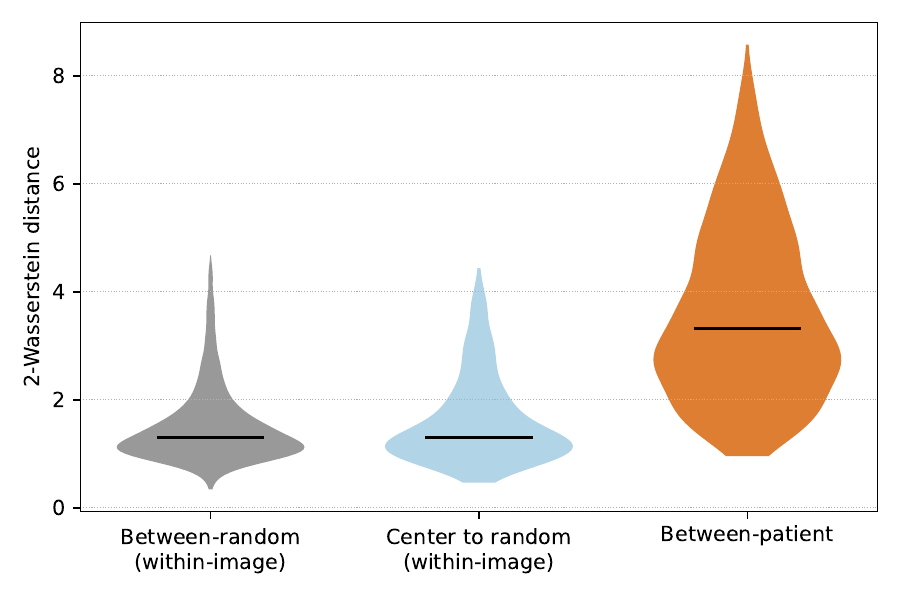}
    \caption{Distribution of $2$-Wasserstein distances for three comparison types: between pairs of random crops of the same image (between-random), between the center crop and random crops of the same image (center to random), and between center crops of different individuals (\textcolor{black}{between-patient}). \textcolor{black}{The first two are within-image comparisons.} Horizontal lines show medians.}
    \label{fig:cropping-distributions}
\end{figure}

In the third experiment, we compared \textcolor{black}{within-image} crop variation to \textcolor{black}{between-patient} $2$-Wasserstein distances from experiment two. The mean \textcolor{black}{between-patient} distance was \textcolor{black}{$3.6054$ (s.d. $0.7334$)}, compared to a mean \textcolor{black}{within-image} distance of \textcolor{black}{$1.6189$ (s.d. $0.5466$)}; see Fig~\ref{fig:cropping-distributions}. A paired $t$-test confirmed that \textcolor{black}{between-patient} distances were significantly larger (\textcolor{black}{$t=-21.543$}, $p<0.0001$), with \textcolor{black}{between-patient} distances approximately twice the magnitude of \textcolor{black}{within-image} variation.

These results validate the analytical approach, that the topological signal distinguishing individuals is substantially larger than the noise introduced by the choice of crop.

\section{Discussion}

\subsection{\textcolor{black}{Cohort selection and handling}}
    We were only able to image a relatively small amount of patient \textcolor{black}{blood draws} for this experimental study. This inevitably limits the conclusions we can draw from the results. However, \textcolor{black}{limiting} our investigations to a smaller cohort enabled us to experiment and develop a method that may be refined, enabling later investigations in larger cohorts. 

    The blood \textcolor{black}{draws} were collected in a reproducible manner, using standard equipment, sample handling and storage. This is important for future applicability of the method, as it may thus easily be applied to other patient populations and settings. Use of freeze-thawed blood plasma enables transport of \textcolor{black}{draws} between laboratories and analyses of stored material, broadening the scope of the method.
    
\subsection{Patient population}   
    Cancer is a major global and personal health burden for afflicted patients \cite{Cancer2024}. After the cancer itself, thrombosis is the most common cause of death in patients \cite{Costamagna2024}. Therefore, understanding and \textcolor{black}{identifying} the coagulation mechanisms behind this increased thrombotic risk is crucial. Patients with \textcolor{black}{esophageal} cancer represent a disease model in this study; however, fibrin network formation should be investigated further in this and other cancer populations in order to draw general conclusions.

    To the authors' knowledge, fibrin network formation has never been examined in a healthy population. This work should be undertaken in order to investigate if a reference range could be calculated, with the goal of developing a \textcolor{black}{standardized} analysis framework, enabling cross-population comparisons in the future.

\subsection{Strengths, limitations and implications for future research}
    To the authors' knowledge, this is the first publication using confocal microscopy to capture fibrin network structures in cancer patients, allowing for a longitudinal analysis of clot architecture over time. Furthermore, this dataset enables a comparative assessment of how different treatment strategies influence fibrin network topology throughout the post-surgical period. By applying TDA, we aimed to quantify structural changes in fibrin networks across \textcolor{black}{draw times} and between treatment groups, providing insights into the dynamic effects of post-surgical therapies on clot stability and thrombotic risk.
 
    \textcolor{black}{In summary, we found no significant differences in fibrin network characteristics in patients who received 10 versus 30 days of anticoagulant prophylaxis following surgery for \textcolor{black}{esophageal} cancer. Similarly, we were not able to identify significant changes in fibrin network formation from before to after surgery.}

    The small size of the data set, and a lack of a \textcolor{black}{\emph{healthy}} baseline to compare with means we are unable to make significant conclusions from this study. We consider it a \emph{proof of concept} that persistent homology \textcolor{black}{can be applied to fibrin networks from a clinical cohort, as the resulting topological summaries are robust to a choice of crop, and distinguish z-stacks containing debris, although they did not distinguish treatment groups or blood draw times in this cohort}. In order to develop a proper framework for using TDA to \textcolor{black}{analyze} these networks, we need to establish an understanding of what a \emph{control} fibrin network should look like topologically. To do this, a study of fibrin networks in plasma from healthy volunteers needs to be conducted. However, this study represents a \textcolor{black}{foundational} step towards a better understanding of the significance of fibrin network topology and its possible clinical significance \textit{in vivo}.

\subsection*{Acknowledgements}
 The authors \textcolor{black}{wish} to thank Dr. Dietmar Fries for graciously hosting TG in his lab and sharing his expertise with confocal imaging. TG and YBB would like to thank Lisbeth Fajstrup and Anne Marie Svane for their insightful comments at the start of this project. We furthermore wish to thank the mathematical research institute MATRIX in Australia where part of this research was conducted, and particularly thank Vanessa Robins, Britt Fasy, Renata Turke\v{s} and Kelly Maggs with whom we discussed this project in-depth as part of the program at MATRIX.
 The work was funded by Aarhus University, Aarhus University Hospital, Oda and Hans Svenningsens' Foundation and Danish Society for Thrombosis and Haemostasis (The Steen Husted Grant), the Danish Cardiovascular Academy and the Austrian Science Fund (FWF) 10.55776/ESP9584724.
 TB acknowledges support by an Australian Government Research Training Program (RTP) Scholarship, and the Danish Data Science Academy (DDSA) research visit Grant ID 2025-5524. Additionally, TR was supported by a DDSA Travel Grant (ID 2024-3690).
 For open access purposes, the author has applied a CC BY public copyright license to any author-accepted manuscript version arising from this submission.


\end{document}